\documentstyle[prb,aps,epsf,twocolumn]{revtex}
\newcommand{\eq}{\begin{equation}}
\newcommand{\ee}{\end{equation}}

\newcommand{\nb}{{\bar{n}}}

\newcommand{\va}{{\vec{a}}}
\newcommand{\vrr}{{\vec{r}}}

\newcommand{\vq}{{\vec{q}}}
\newcommand{\vs}{{\vec{s}}}
\newcommand{\vR}{{\vec{R}}}
\newcommand{\vS}{{\vec{S}}}
\newcommand{\vQ}{{\vec{Q}}}
\newcommand{\vX}{{\vec{X}}}
\newcommand{\vPi}{{\vec{\Pi}}}

\def\cH{{\cal H}}
\def\ddt{{\partial\over \partial t}}
\def\lam{\lambda}
\def\t{\theta}

\def\a{\alpha}
\def\b{\beta}
\def\e{\epsilon}
\def\o{{\omega}}

\def\dd{d^{\dagger}}

\def\half{{1\over2}}
\def\third{{1\over3}}
\def\twof{{2\over5}}

\def\rhob{{\bar \rho}}
\def\intq{{\int {d^2q\over(2\pi)^2}}}
\def\ints{{\int {d^2s\over(2\pi)^2}}}

\def\hz{{\hat z}}

\def\eqa{\begin{eqnarray}}
\def\eea{\end{eqnarray}}
\def\prl{{Phys. Rev. Lett.}}
\def\prb{{Phys. Rev. {\bf B}}}
\def\jpc{{Jour. Phys. {\bf C}}}

\begin{document}
\draft
\flushbottom
\twocolumn[
\hsize\textwidth\columnwidth\hsize\csname @twocolumnfalse\endcsname
\title{ Hamiltonian Theory of the FQHE: Conserving Approximation for Incompressible Fractions }
\author{  Ganpathy Murthy}
\address{
Department of Physics and Astronomy,  University of Kentucky, Lexington KY 40506\\
}
\date{\today}
\maketitle
\tightenlines
\widetext
\advance\leftskip by 57pt
\advance\rightskip by 57pt

\begin{abstract}
A microscopic Hamiltonian theory of the FQHE developed by Shankar and
the present author based on the fermionic Chern-Simons approach has
recently been quite successful in calculating gaps and finite
tempertature properties in Fractional Quantum Hall states. Initially
proposed as a small-$q$ theory, it was subsequently extended by
Shankar to form an algebraically consistent theory for all $q$ in the
lowest Landau level. Such a theory is amenable to a conserving
approximation in which the constraints have vanishing correlators and
decouple from physical response functions. Properties of the
incompressible fractions are explored in this conserving
approximation, including the magnetoexciton dispersions and the
evolution of the small-$q$ structure factor as $\nu\to\half$. Finally,
a formalism capable of dealing with a nonuniform ground state charge
density is developed and used to show how the correct fractional value
of the quasiparticle charge emerges from the theory.
\end{abstract}
\vskip 1cm

]
\narrowtext
\tightenlines
\section{Introduction}
\label{sec1}

The Fractional Quantum Hall (FQH) effect\cite{fqhe-ex} has introduced
us to new states of electrons in high magnetic
fields\cite{perspectives}. Laughlin\cite{laugh} explained the essence
of the effect for fractions $\nu=1/(2p+1)$ and showed that the
electrons form  strongly correlated states which are incompressible
and have quasiparticle excitations with fractional charge\cite{laugh}
$e^*=e/(2p+1)$.

A unified understanding of all fractions $\nu=p/(2sp+1)$ was achieved
by the Composite Fermion picture of Jain\cite{jain-cf}. In this
picture, the true quasiparticles are electrons dressed by $2s$ units
of statistical flux, which are called Composite Fermions (CFs). At a
mean field level, the CFs then see a reduced field $B^*=B/(2sp+1)$, in
which they fill $p$ CF-Landau levels (CF-LLs), and exhibit the integer
QHE. This picture has been very successful in describing the FQH
states\cite{jain-cf-review}.

Numerical diagonalizations of small systems\cite{hal-rez,exact-ex} or
the use of trial wavefunctions\cite{jain-cf-review,jain-ex} are the
best approaches for calculating many properties of FQH systems such as
ground state energies, gaps, magnetoexciton dispersions, ground state
spin transitions, etc. However, there are other interesting properties
which are not accessible to these approaches, the most important of
which are dynamical response functions. Partly to address  these
issues, and also to gain a better understanding of FQH states, several
field-theoretic approaches were developed during the past fifteen
years. Most of these approaches are based on the Chern-Simons(CS)
transformation\cite{cs-trans}, which performs flux attachment via the
CS gauge field to obtain either bosons\cite{gcs,gmcs,zhk,read1} or
fermions\cite{lopez}. These theories have provided us with a link
between the microscopic formulation of the problem and experiment,
both for incompressible and compressible states\cite{kalmeyer,hlr}.

Motivated by some outstanding issues in previous work\cite{ssh}, Shankar and the
present author recently developed a hamiltonian CS theory for the FQH
states
\cite{us1,us2}. Inspired by  the work
of Bohm and Pines\cite{bohm-pines} on the 3D electron gas, we enlarged
the Hilbert space to introduce $n$ high-energy magnetoplasmons degrees
of freedom, ($n$ also being the number of electrons) at the same time
imposing an equal number of constraints on physical states.  Upon
ignoring the coupling between the oscillators and the fermions we
obtained some well known wavefunctions
\cite{laugh,jain-cf}. A final
canonical transformation was employed to decouple the fermions from
the oscillators  in the infrared limit.

After the final canonical transformation, we obtained the following
expressions\cite{us2} for the electronic charge density $\rho_e$ and the
constraint $\chi$ for $\nu=p/2p+1$ at small $q$:
\eqa
&\rho_e(\vq)={q \over \sqrt{8\pi}}c (A(\vq)+
A^{\dagger}
(-\vq))\nonumber\\
&+\sum\limits_{j} e^{-i\vq\cdot\vrr_j} -{il_{}^{2}\over1+c }  \sum\limits_{j}\vq \times
\vPi_je^{-i\vq\cdot\vrr_j}
\label{rhoesmall}\\
&\chi(\vq)=\sum\limits_{j} e^{-i\vq\cdot\vrr_j} +{il_{}^{2}\over c(1+c) }  \sum\limits_{j} \vq \times
\vPi_je^{-i\vq\cdot\vrr_j}
\label{rhovsmall}
\eea
where $A,A^{\dagger}$ refer to the annihilation and creation operators
of the magnetoplasmon oscillators, $l =1/\sqrt{eB}$ is the magnetic
length, ${\vec\Pi}_j={\vec P}_j+e{\vec A}^*(r_j)$ is the velocity
operator of the CFs, and $c=\sqrt{2\nu}$. Instead of using the
expression of eq(\ref{rhoesmall}) for the electronic charge density we
added a multiple of the constraint to it in subsequent
calculations. The rationale for doing so is that in an exact
calculation in the physical subspace (which is annihilated by the
constraints) using the operator $\rho_e-\xi \chi$ with {\it arbitrary}
$\xi$ instead of $\rho_e$ should not change the physical
answer. However, in an approximate calculation in the full space
different choices of $\xi$ will produce different answers. The choice
$\xi=c^2=2\nu$ is particularly advantageous, as will be seen
below. With this ``preferred'' choice, the expression for the density
becomes
\eqa
\rho_p(\vq)={q \over \sqrt{8\pi}}c (A(\vq)+
A^{\dagger}
(-\vq))&\nonumber\\
+{1\over2p+1}\sum\limits_{j} e^{-i\vq\cdot\vrr_j} -{il_{}^{2}}  \sum\limits_{j} \vq \times
\vPi_je^{-i\vq\cdot\vrr_j}&
\label{rhoesmallpref}
\eea

The oscillator piece saturates Kohn's
theorem\cite{kohn}. The rest, to be called $\rhob_p$, has the following 
properties:

\begin{itemize}
\item{} The presence of the velocity operator $\vPi$ in $\rhob_p$ 
shows that the CFs respond to the effective field $B^*$.
\item{} $\rhob_p$ satisfies the
magnetic translation algebra (MTA)\cite{GMP} to lowest leading
order. Since this is the algebra of the electron density in the lowest
Landau level (LLL), we have performed the LLL projection correctly in
the infrared.
\item{} Note that $\rhob_p$ is a sum of a monopole with charge
$e^*=e/(2p+1)$, which is the charge associated with the CF, and a
dipole piece\cite{read2} which alone survives at $\nu=1/2$.  (A number
of recent constructions have emphasized this dipolar
aspect\cite{dh,pasquier,read3,stern}).
\item{} We also find
that that as $\vq\to 0$ all transition matrix elements of $\rhob_p$
from the HF ground state vanish at least as $q^2$ (a consequence of
Kohn's theorem\cite{kohn} for incompressible states\cite{GMP}).
\end{itemize}

These are all nonperturbatively correct properties of the true
electronic density operator in the LLL. In our formulation the full
Hilbert space of the CFs is too large, and the physical LLL will be
reached only after the constraints have been properly imposed.
Unfortunately, we do not know how to select the physical states
operationally in the final representation. The use of $\rhob_p$
represents a shortcut that obviates the need to impose constraints.
The idea is that since $\rhob_p$ displays the same properties in the
full space that the true electronic density is known to possess in the
physical space, calculations carried out with $\rhob_p$ in the full
space will hopefully provide a reasonable approximation to the true
physical answers.

The Hamiltonian of the low-energy sector now is
\eq
H=\half \int {d^2 q\over(2\pi)^2} v(q) \rhob_p(-q) \rhob_p(q)\label{ham}.
\ee
where $v(q)$ is the electron-electron interaction.  

Once the constraints have been used to construct $\rhob_p$, they are 
ignored in subsequent calculations. One assumes
that the Hartree-Fock(HF) approximation is sufficient to capture the
correct physics. This assumption is borne out by calculations of
gaps\cite{single-part,scaling} and the temperature dependence of the
spin polarization\cite{pvst,full-disc}. It even provides a reasonable
qualitative description of the magnetoexciton dispersions\cite{me-us}.

Despite these successes, the above formulation has a few
disadvantages. Firstly, the expressions for the density and constraint
operators are known only to first order in $q$. The use of the
small-$q$ preferred density operator for all $q$ renders the gap
infinite for the Coulomb interaction, and one has to use a potential
that is exponentially suppressed at large $q$ in order to get finite
results\cite{thick1,thick2}. In computing magnetoexciton dispersions
using the small-$q$ approach\cite{me-us} the author used a cutoff in
the number of CF-Landau levels (CF-LLs) corresponding to keeping the
correct number of single-particle states in the Hilbert space to
obtain finite results for the magentoexciton dispersions for any
interaction. However, for realistic potentials the dispersions were
still strongly dependent on the CF-LL cutoff\cite{me-us}, casting
doubt on the robustness of the approach. A related disadvantage is
that the density operator satisfies the MTA only to leading order in
$q$.

As stated above, the preferred density operator was obtained
by adding a multiple of the constraint\cite{us2} to the canonical
transform of the original electron density. This made certain
nonperturbative features of the charge density explicit. However, it
would be desirable and more satisfying to see these nonperturbative
features emerging from the formalism rather than having to put them
in.

Shankar\cite{aftermath} extended the previous small-$q$ approach to
all orders in $q$ by exponentiating the first two terms of the
small-$q$ expressions for the density (eq(\ref{rhoesmall})) and the
constraint (eq(\ref{rhovsmall})). The resulting expressions
(considered in more detail in the next section) have some very
desirable algebraic properties; the density operator obeys the
magnetic translation algebra exactly at all $q$, and the density and
the constraint commute. This implies that the Hamiltonian constructed
from the density operator commutes with the constraint, which makes a
standard conserving approximation\cite{kada-baym} possible.  In such
an approximation, the correlator of the constraint with either the
physical density or itself is explicitly zero. It is worth noting that
the algebra found by Shankar reduces to that found by Pasquier and
Haldane\cite{pasquier} for bosons at $\nu=1$ (in this case $c^2=\nu$
in the above formulas, and bosons at $\nu=1$ are mapped to fermions in
zero average field, just as electrons are at $\nu=\half$). This case
was treated in great detail by Read\cite{read3}, who computed bosonic
response functions in a conserving approximation.

In this paper, we will follow this conserving route to calculating
response functions for the incompressible fractions. This corresponds
to using the HF approximation on the Hamiltonian to calculate
single-particle energies and calculating bosonic correlators in the
time-dependent HF (TDHF) approximation. Several results of interest
are obtained. Firstly, it is known that the spectral function of the
correlator of the projected electron density operator cannot contain
terms of order $q^2$, since this would violate Kohn's
Theorem\cite{kohn}. Naively, a density perturbation produces excitons
from filled states to empty states. The electronic density operator
has matrix elements of order $q$ to some of these states, and it is
interesting to see how these disappear in the final correlator. It
turns out that there is always one linear combination of naive
excitons with a zero eigenvalue in the TDHF approximation. This
corresponds in a very close sense (to be made precise later) to the
constraint, and represents the unphysical sector of the theory. The
electronic density operator does not couple to this unphysical sector
in the TDHF approximation, and does indeed have matrix elements of
order $q^2$ or higher to the physical excitonic states, in compliance
with Kohn's theorem. As a consequence of the $q^2$ matrix elements the
structure factor $S(q)$, which is the equal time correlator
$<\rho_e(q,t)\rho_e(-q,t)>$, should go as $q^4$ in the incompressible
states. However, Read's work\cite{read3} shows that at $\nu=\half$
this behaves as $q^3\log{q}$. We will show that the for the fraction
$\nu=p/(2p+1)$, the coefficient of the $q^4$ term in the structure
factor diverges (for $p>1$) as
\eq
{1\over2}{p^4-3p^3+{5\over4}p^2+3p+{7\over4}\over p^2-1}
\label{sfp}\ee
This result shows that as $p\to\infty$, a power series expansion of
$S(q)$ must fail, and is consistent with the above nonanalytic
result\cite{read3}.

We will also present results for the magnetoexciton dispersions for
$1/3$ and $2/5$. The positions of the minima are found to be exactly
at the values of $ql_0$ found
numerically\cite{jain-cf-review,hal-rez,exact-ex,jain-ex}. This
attests to the fact that although the all-$q$ formalism was derived by
extending a small-$q$ approach, it nevertheless manages to capture the
structure of the magnetoexcitons up to $ql_0\approx 1.5$.

The fractional quasiparticle charge\cite{laugh,frac-charge,jain-count} is one of
the most striking and robust features of the FQHE, and it must emerge
from any theory that correctly describes the FQHE. To see how this
appears in the all-$q$ Hamiltonian theory, a further extention of the
formalism is necessary. Shankar's all-$q$ formalism\cite{aftermath} is
designed to work best in a uniform background. In an inhomogeneous
background there will be a nonuniform average of the electronic charge
density $n(\vrr)=\nb+\delta n(\vrr)$, which we expect by the
principles of flux attachment to lead to a nonuniform effective
magnetic field. Clearly, adding a localized quasiparticle produces
just such a nonuniform charge density and effective field. A version of the
all-$q$ theory applicable to first order in $\delta n/\nb$ will be
presented, and used to show the emergence of the fractional
quasiparticle charge. Such a formalism is also useful if one wants to treat
disorder or edges in a conserving approximation.

The structure of this paper is as follows: In Section\ \ref{sec2} we
provide a brief introduction to the Hamiltonian theory of the FQHE,
including the single-particle states. In Section\ \ref{sec3} we will
present the structure of the conserving approximation and the details
of how it is carried out. In Section\ \ref{sec4} we will analyse the
TDHF matrix, show that it has zero modes corresponding to the
unphysical modes.  The result for the small-$q$ structure factor for
$\nu=p/(2p+1)$ will be derived in Section\ \ref{sec5}. In Section\
\ref{sec6} we will present results for the magnetoexciton
dispersions for $\third$ and $\twof$ for $q$ not too small. In
Section\ \ref{sec7} we will present the extention of the all-$q$
formalism to the case of nonuniform density, and see how the
fractional quasiparticle charge emerges naturally in the theory. We
end with some conclusions. 

\bigskip
\section{Hamiltonian theory of the FQHE}
\label{sec2}

As described in the introduction, the small-$q$ version of the
Hamiltonian theory\cite{us1,us2} of the FQHE was derived by introducing extra
degrees of freedom which assumed the role of the magnetoplasmons, and
supplementing the Hamitonian with constraints. After decoupling this
led to a theory with many desirable properties, but also some
deficiencies, notably involving the consistency of the theory for all
$q$.

We will take the all-$q$ extention of this formalism as described by
Shankar\cite{aftermath} as our starting point. In this formalism two
sets of guiding center coordinates are associated with each CF.

\eqa
\vR_e=&\vrr-{l^2\over1+c}\hz\times\vPi\label{redef}\\
\vR_v=&\vrr+{l^2\over c(1+c)}\hz\times\vPi\label{rvdef}
\eea
Here $c^2=2\nu$, and $\vrr$ and $\vPi$ are the position and velocity
operators of the CF. The algebra of these objects is of fundamental
importance.
\eqa
\left[R_{e,\a},R_{e,\b}\right]=& -il^2\e_{\a\b}\label{crrere} \\
\left[R_{e,\a},R_{v,\b}\right]=& 0 \label{crrerv}\\
\left[R_{v,\a},R_{v,\b}\right]=& i{l^2\over c^2}\e_{\a\b} \label{crrvrv}
\eea

The first set of coordinates $\vR_e$ is identified with the guiding
center coordinates of the electron, and the second set $\vR_v$
corresponds to the guiding center coordinates of an object with a
magnetic algebra charge of $-e/c^2$. We will call this object the {\it
pseudovortex} (for brevity, Shankar simply calls it the
vortex\cite{aftermath,full-disc}). Upon projection to the physical
subspace we expect that this coordinate will disappear, leaving behind
the proper correlations between electrons, that is, vortices. Magnetic
translation operators $T_{e}$, $T_{v}$ and densities $\rho_e$,
$\rho_v$ can be formed from the two sets

\eqa
\rho_e(\vq)=&e^{-q^2l^2/4} T_e(\vq)=e^{-q^2l^2/4}\sum\limits_i e^{-i\vq\cdot\vR_{e,i}}\label{eldensity}\\
\rho_v(\vq)=&e^{-q^2l^2/4c^2} T_v(\vq)=e^{-q^2l^2/4c^2}\sum\limits_i e^{-i\vq\cdot\vR_{v,i}}\label{vordensity}
\eea

The operator $T_e$ obeys the algebra of the electronic magnetic translation operators 
projected to the LLL\cite{GMP}
\eq
[T_e(\vq_1),T_e(\vq_2)]=2i\sin({l^2\over2} \vq_1\times\vq_2)T_e(\vq_1+\vq_2)
\label{tete}\ee
while the operator $T_v$ satisfies a similar algebra 
\eq
[T_v(\vq_1),T_v(\vq_2)]=-2i\sin({l^2\over2c^2} \vq_1\times\vq_2)T_v(\vq_1+\vq_2)
\label{tvtv}\ee

The density formed from the electronic guiding center is identified
with the electronic density operator, while that formed from the
pseudovortex guiding center is the constraint\cite{aftermath}. It is
easy to verify that the small-$q$ limit of these expressions indeed
reduce to those derived in the small-$q$ theory
(eqs(\ref{rhoesmall},\ref{rhovsmall})). Since they commute with each
other, the Hamiltonian formed from $\rho_e$ alone will commute with
the constraint.

In previous uses of this all-$q$ formalism\cite{full-disc}, the {\it
preferred} density $\rho_{p}=\rho_e-c^2\rho_v$ was used. The preferred
density has many desirable features, as described in the
introduction, and gives numbers in good agreement with numerical and
experimental results for many physical
properties\cite{single-part,pvst,full-disc}. However, in addressing
issues of principle, such as the form of the density correlator at
small-$q$ for $\nu=1/2$\cite{comment,stern}, naive calculations based
on the preferred density fail to produce the correct answer, and a
conserving calculation\cite{kada-baym,read3} is needed.

Before we can compute density correlators, we need to
construct the single-particle states of the theory. Our Hamiltonian is
\eq
H=\intq v(q) \rho_e(\vq)\rho_e(-\vq)
\label{hame}\ee
From eq(\ref{redef}) we see
that the coordinate $R_e$ contains $\vec{\Pi}$, the velocity operator
in the effective field. Thus, Landau level states in the effective
field make a natural basis in which to represent the various
operators. In the Landau gauge, these states are
\eq
\phi_{n,X}=Ce^{iXy/(l^*)^2} e^{-(x-X)^2/2(l^*)^2}H_n((x-X)/l^*)
\ee
where $C$ is a normalization constant, $H_n$ are the Hermite
polynomials, $l^*=1/\sqrt{eB^*}$ is the magnetic length in the
effective field, $n$ represents the Composite Fermion-Landau Level
(CF-LL) index, and $X=2\pi j (l^*)^2/L$ is the degeneracy index (the
degeneracy of each CF-LL is ${L^2\over2\pi(l^*)^2}$). In terms of
these states the translation operators can be
represented as follows:
\eqa
&T_e(\vq)=\sum\limits_{\{n_i\}X}e^{-iq_xX}\dd_{n_1,X-{Q_yl^*\over2}} d_{n_2,X+{Q_yl^*\over2}} \rho_{n_1n_2}(\vq)\\
&T_v(\vq)=\sum\limits_{\{n_i\}X}e^{-iq_xX}\dd_{n_1,X-{Q_yl^*\over2}} d_{n_2,X+{Q_yl^*\over2}} \chi_{n_1n_2}(\vq)
\eea
where the matrix elements are
\eqa
&\rho_{n_1n_2}(\vq)=(-1)^{n_<+n_2}\sqrt{{n_<!\over n_>!}}e^{i(n_1-n_2)(\t_q-\pi/2)}\nonumber\\
& (cQ/\sqrt{2})^{|n_1-n_2|}e^{-c^2Q^2/4}L_{n_<}^{|n_1-n_2|}(c^2Q^2/2)\label{rhoe12}\\
&\chi_{\nu}(\vq)=(-1)^{n_<+n_2}\sqrt{{n_<!\over n_>!}}e^{i(n_1-n_2)(\t_q-\pi/2)}\nonumber\\
& (Q/c\sqrt{2})^{|n_1-n_2|}e^{-Q^2/4c^2}L_{n_<}^{|n_1-n_2|}(Q^2/2c^2)\label{rhov12}
\eea
Here $c=\sqrt{2\nu}$, $Q=ql^*$, $n_<$ ($n_>$) is the lesser (greater)
of $n_1,\ n_2$, and $L_n^k$ are the Laguerre polynomials.

It can be shown that the CF-LL states are in fact Hartree-Fock (HF)
states of the Hamiltonian\cite{me-us,full-disc}, that is, they are the
best possible Slater determinantal states for this problem. The
single-particle HF energies are
\eq
\e(n)=\half\intq v(q)e^{-q^2l^2/2}\sum_{m} (1-2N_F(m))|\rho_{nm}(\vq)|^2
\label{hfenergy}\ee
Note that the energy depends only on the CF-LL index, and not on the
degeneracy index. The first term corresponds to the Hartree energy,
while the second term is the Fock energy. When the preferred density
is used the Hartree contribution is nontrivial, and depends on
$n$. However, when the electron density of eq(\ref{eldensity}) is
used, as we do here, the Hartree energy is a constant independent of
$n$ (the methods of the Appendix can be adapted to show this).

\section{The Conserving Approximation}
\label{sec3}

To illustrate the idea of a conserving approximation\cite{kada-baym},
let us go back to the Hamiltonian constructed from the all-$q$ density
operator $\rho_e(\vq)$, eq(\ref{hame}).  This should be supplemented by
constraints that demand that the pseudovortex density should
annihilate all physical states
\eq
\rho_v(\vq)|\Psi_{phys}>=0
\ee 

Now define a time-ordered  Green's function $G_{ve}(q,t)$ as
follows (one can similarly define $G_{ee}$ and  $G_{vv}$)
\eqa
G_{ve}(\vq,t-t')=&-i<T\rho_v(\vq,t)\rho_e(-\vq,t')>\nonumber\\
=&-i\Theta(t-t')<\rho_v(\vq,t)\rho_e(-\vq,t')>\nonumber\\
&-i\Theta(t'-t)<\rho_e(-\vq,t')\rho_v(\vq,t)>
\eea
where $T$ is the time ordering operator. Consider the time development
of this Green's function
\eqa
-i\ddt G_{ve}&(\vq,t-t')=-\delta(t-t')<[\rho_v(\vq,t),\rho_e(-\vq,t')]>\nonumber\\
&-i<T[H,\rho_v(\vq,t)]\rho_e(-\vq,t')>
\eea
Since $\rho_v$ and $\rho_e$ commute, $\rho_v$ also commutes with
$H$. One immediately sees that $G_{ve}$ is a constant. If one sets the
constraint to zero initially, then it remains zero, and all its
correlators also remain zero.

The above is true as an exact statement. However, in practice one uses
some approximation scheme to calculate the Green's function. In our
case, an approximation that respects $G_{ve}=G_{vv}=0$ will be 
conserving. Let us see what a natural approximation scheme might
be. Consider the time-ordered Green's function $G_{ee}(\vq,t)$ defined
as
\eq
G_{ee}(\vq,t-t')=-i<T\rho_e(\vq,t)\rho_e(-\vq,t')>
\ee
and look at its time development. We have
\eqa
i\ddt G_{ee}(\vq,t)=&\delta(t)<[\rho_e(\vq,t),\rho_e(-\vq,0)]>\nonumber\\
&-i<T[H,\rho_e(\vq,t)]\rho_e(-\vq,0)>
\eea
From the commutation relations of $\rho_e$ with itself
(eq(\ref{tete})), it can be seen that a Green's function involving
three densities will arise. Further time derivatives will involve
higher-order density correlators. The natural way to truncate this
hierarchy is to make a mean-field approximation at some stage that
reduces a product of densities to a single density. One of the
simplest of such approximations\cite{kada-baym} reduces $[H,\rho_e]$,
which is a product of four fermi operators, to a product of only two,
by using the averages
\eqa
\dd_{\a_1}\dd_{\a_2}&d_{\b_2}d_{\b_1}\to <\dd_{\a_1}d_{\b_1}>\dd_{\a_2}d_{\b_2} 
+<\dd_{\a_2}d_{\b_2}>\dd_{\a_1}d_{\b_1} \nonumber\\
&-<\dd_{\a_1}d_{\b_2}>\dd_{\a_2}d_{\b_1}
-<\dd_{\a_2}d_{\b_1}>\dd_{\a_1}d_{\b_2}
\eea
Here $<\dd_{\a}d_{\b}>=\delta_{\a\b}N_F(\a)$, where $N_F(\a)$ is the
Fermi occupation of the single-particle state $\a$.  

Using the HF states and their occupations in the above truncation is
known as the time-dependent HF (TDHF) approximation. In the electron
gas it is known that this approximation is conserving (in that case
the conservation laws to be satisfied result from gauge
invariance\cite{kada-baym}). Read\cite{read3} has shown that this
approximation is conserving for the $\nu=1$ boson problem as well. We
will explicitly show below that TDHF is conserving for all principal
fractions.

Let us proceed to some more explicit details.

\subsection{TDHF Formalism}
\label{sec3.1}

The final all-$q$ theory has mapped the original strongly correlated
electron problem into a weakly correlated CF problem with
constraints. This CF problem has LL structure in the effective field,
and integer filling for CFs. All the techniques that have been brought
to bear on the problem of computing magnetoexcitons in the integer
quantum Hall (IQH) states can be applied to this problem. The
computation of magnetoexciton dispersions has a long
history\cite{me-history} in the IQHE. Many different
approximations\cite{me-history,kallin} have been employed to obtain
these dispersions in the IQHE, most of which can be subsumed into a
unified TDHF treatment\cite{macd1,macd2}. In this approach one allows
the hamiltonian to scatter particle-hole excitations with different
Landau-level indices into each other, and diagonalizes the resulting
matrix. As mentioned before, this method has been applied to the FQHE
using the small-$q$ hamiltonian approach\cite{me-us}. Most of the
details can be found there, so the following will touch on the
highlights of the formalism.

The TDHF matrix can be derived  from many different points of
view. We will find it most convenient to follow the equation of motion
approach of Rickayzen\cite{operator-rpa}. We start with the
magnetoexciton operator
\eq
O_{m_1m_2}(\vq)=\sum_{X} e^{-iq_xX} \dd_{m_1,X-{Q_yl^*\over2}} d_{m_2,X+{Q_yl^*\over2}}
\ee

This represents a CF-particle-hole excitation with a definite momentum
$\vq$. To avoid a proliferation of indices we will use the shorthand
$\mu={m_1m_2}$ (and sometimes $\nu={n_1n_2}$, not to be confused with the
filling factor!).  Now we define a time-ordered Green's function of
two of these operators
\eq
G(\mu;\mu';\vq;t-t')=-i<TO_{\mu}(\vq,t)O_{\mu'}(-\vq,t')>
\ee
Taking the time derivative we obtain
\eqa
&-i\ddt G(\mu;\mu';\vq;t)=-i<T[H,O_{\mu}(\vq,t)]O_{\mu'}(-\vq,0)>\nonumber\\
&-\delta(t)<[O_{\mu}(\vq,t),O_{\mu'}(-\vq,0)]>
\label{eqofmotion}\eea
Now one Fourier transforms $G$ to convert it to a function of
frequency $\o$ according to 
\eq
G(t)=\int\limits_{-\infty}^{\infty} {d\o\over2\pi} e^{i\o t} G(\o)
\ee
The average value of the equal time commutator in the ground state
(the final term of eq(\ref{eqofmotion})) can be carried out to give
\eqa
<\left[O_{\mu}(\vq),O_{\mu'}(-\vq)\right]>=&{L^2\over2\pi(l^*)^2} \delta_{m_1m_2'} \delta_{m_2m_1'} \times\nonumber\\
&(N_F(m_1)-N_F(m_2))
\eea
Now the Hamiltonian is expressed as a four-fermi operator
\eqa
&H={1\over 2L^2}\sum\limits_{\vq,\{X_i\}\{n_i\}} v(q) e^{-iq_x(X_1-X_2)} \rho_{n_1n_2}(\vq) \rho_{n_3n_4}(-\vq)\nonumber\\
& \dd_{n_1,X_1-{Q_yl^*\over2}}d_{n_2,X_1+{Q_yl^*\over2}} \dd_{n_3,X_2+{Q_yl^*\over2}}d_{n_4,X_2-{Q_yl^*\over2}}
\eea
One performs the commutator $[H,O_{\mu}]$, and in the TDHF
approximation one contracts the resulting four-fermi operator to give
a two-fermi operator. After the sums over degeneracy indices are
performed one finds
\eqa
&\left[H,O_{\mu}(\vq)\right]=(\e(m_1)-\e(m_2))O_{\mu}(\vq)+\nonumber\\
&(N_F(m_2)-N_F(m_1))\sum\limits_{\nu} O_{\nu}(\vq) 
\bigg({v(q)\over 2\pi(l^*)^2} e^{-q^2l^2/2}\rho_{n_1n_2}(\vq)\times\nonumber\\
& \rho_{m_2m_1}(-\vq)
-\ints v(s) e^{-s^2l^2/2}\rho_{n_1m_1}(\vs) \rho_{m_2n_2}(-\vs) e^{i\vS\times\vQ}\bigg)
\label{tdhfeq}\eea
One can now define a vector $\Psi(\mu)$, which corresponds to the operator
\eq
O_{\Psi}=\sum\limits_{\mu} \Psi(\mu) O_{\mu}
\ee
Thus the action of commuting with $H$ on $O_{\mu}$ in the TDHF
approximation can be represented as the right-multiplication of $\Psi$
of a matrix $\cH(\mu;\nu;\vq)$ in the space of these vectors
\eqa
&\cH(\mu;\nu;\vq)=\delta_{m_1n_1}\delta_{m_2n_2} (\e(m_1)-\e(m_2))+\nonumber\\
&(N_F(m_2)-N_F(m_1))
\bigg({v(q)\over 2\pi(l^*)^2} e^{-q^2l^2/2}\rho_{n_1n_2}(\vq) \rho_{m_2m_1}(-\vq)\nonumber\\
-&\ints v(s)e^{-s^2l^2/2} \rho_{n_1m_1}(\vs) \rho_{m_2n_2}(-\vs) e^{i\vS\times\vQ}\bigg)
\label{tdhfmat}\eea
It is clear that finding the eigenvalues and eigenvectors of this
matrix will also enable us to solve for the Green's function. Assume
that one has found the right and left eigenvectors and corresponding
eigenvalues, labelled by $\a$
\eqa
\cH(\mu;\nu;\vq)\Psi_{\a}^{R}(\nu;\vq)=&E_{\a}(q)\Psi_{\a}^{R}(\nu;\vq)
\label{psir}\\
\Psi_{\a}^{L}(\mu;\vq)\cH(\mu,\nu;\vq)=&E_{\a}(q)\Psi_{\a}^{L}(\nu;\vq)
\label{psileft}\eea
where sums over repeated indices are implicit. 

We can choose to normalize the eigenvectors in the conventional way
\eq
\Psi_{\a}^{L}(\mu;\vq)\Psi_{\b}^{R}(\mu;\vq)=\delta_{\a\b}
\label{norm}\ee

If the matrix has a complete set of eigenvectors (we will
see in the next subsection that this is not true, and see how to deal
with this situation) then one can expand $(\o-\cH)^{-1}$ as 
\eq
(\o-\cH)^{-1}(\mu;\nu;\vq)=\sum_{\a} \Psi_{\a}^{R}(\mu;\vq){1\over \o-E_{\a}} \Psi_{\a}^{L}(\nu;\vq)
\ee
Now the solution to the Green's function can be expressed as
\eqa
&G_(\mu;\mu';\vq;\o)={L^2\over2\pi(l^*)^2}\sum\limits_{\a}\nonumber\\
& \Psi_{\a}^R(\mu;\vq){1\over\o-E_{\a}}\Psi_{\a}^L(m_2'm_1';\vq)(N_F(m_1')-N_F(m_2'))
\label{specrep}\eea

\section{Formal Analysis of the TDHF Matrix}
\label{sec4}

In this section we will first show that there is a definite connection
between the right and left eigenvectors, and that the eigenvalues of
$\cH$ always come in pairs $\pm E_{\a}$. Next we will show that there
is always one eigenvector of $\cH$ with a zero eigenvalue, and that
this corresponds to the constraint. There is also another zero
eigenvalue which does not correspond to any eigenvector, and the
matrix $\cH$ is {\it defective}, that is, does not have a complete set
of states. We will see that we can nevertheless compute all the
physical Green's functions, since they do not couple to the unphysical
sector.

\subsection{Two Useful Properties of $\cH$}
\label{sec4.1}

It is easy to see that $\cH$ is neither a symmetric not a Hermitian
matrix, so in general one should expect right and left eigenvectors
which are not Hermitian adjoints of each other. However, there is a
definite connection between the right and left eigenvectors with the
same eigenvalue:
\eq
\Psi_{\a}^{R}(\nu;\vq)\propto (N_F(n_2)-N_F(n_1))\big(\Psi_{\a}^{L}(\nu;\vq)\big)^*
\ee
This can easily be verified by noting that
$\rho_{n_1n_2}(\vq)=(\rho_{n_2n_1}(-\vq))^*$, and taking the complex
conjugate of the eigenvalue equation for the left eigenvector
(eq(\ref{psileft})). In fact, any multiple of the above will form a
good right eigenvector. However, to satisfy the normalization
condition eq(\ref{norm}), we need to take
\eq
\Psi_{\a}^{R}(\nu;\vq)=sgn(E_{\a}) (N_F(n_2)-N_F(n_1))\big(\Psi_{\a}^{L}(\nu;\vq)\big)^*
\label{psiright}\ee

The second property is that eigenvalues occur in equal and opposite
pairs $\pm E_{\a}$.  By focusing on the direct term, it is easy to
verify that the phase of a term in $\cH$ is
\eqa
&{\cH_d(\mu;\nu;\vq)\over |\cH_d(\mu;\nu;\vq)|}= (-1)^{n_<+m_<+m_1+n_1} \nonumber\\
&e^{i(\t_q-\pi/2)(n_1+m_2-n_2-m_1)}
(N_F(m_2)-N_F(m_1))
\eea
It is clear from eq(\ref{tdhfmat}) that $\mu$ such that $N_F(m_2)=N_F(m_1)$ are
not coupled to any other modes, and therefore have a time-evolution
given solely by their energy $\e(m_1)-\e(m_2)$. The nontrivial part of
$\cH$ consists of those vectors for which $N_F(m_2)-N_F(m_1)\ne 0$. We
can divide this subspace into a positive energy part with
$N_F(m_2)-N_F(m_1)>0$, and a negative energy part with
$N_F(m_2)-N_F(m_1)<0$. Clearly, for the trivial part of the spectrum,
one can immediately construct a negative energy mode $O_{m_2m_1}$ for
every positive energy mode $O_{m_1m_2}$.

Before we proceed, let us note that rotational invariance guarantees
that the eigenvalues $E_\a(q)$ depend only on the magnitude of $\vq$. In
order to analyse the structure of $\cH$ the choice $\t_q=0$ is
particularly convenient. With this choice, it is easy to see that
\eq
\cH_d(m_1m_2;n_1n_2;q_x)=-\cH_d(m_2m_1;n_2n_1;q_x)
\ee
With a little more effort it is possible to show that the same
relation holds for the exchange part, and thus for the entire matrix
$\cH$, which has the following structure 
\eq
\cH=\left( \begin{array}{cc}
            \cH_{++}&\cH_{+-}\\
            -\cH_{+-}&-\cH_{++}
           \end{array}\right)
\ee
It follows that if 
\eq
(u_+\ u_-)\cH=E(u_+\ u_-)
\label{pm1}\ee
is a left eigenvector of $\cH$ with eigenvalue $E$, then
\eq
(u_-\ u_+)\cH=-E(u_-\ u_+)
\label{pm2}\ee
is a left eigenvector with eigenvalue $-E$. The only exception
that could occur is if $E=0$, in which case $(u_-\ u_+)$ might
possibly be the same as $(u_+\ u_-)$, and one is left with only a
single eigenvector. This case is realized in our conserving
approximation, and we now turn to its analysis.

\subsection{Unphysical States in $\cH$}
\label{sec4.2}

The most important property of $\cH$ is that it always has at least
one left eigenvector with zero eigenvalue, namely
\eq
\Psi_{0}^{L}(m_1m_2;\vq)=\chi_{m_1m_2}(\vq)
\ee

The requisite formalism to show this explicitly is detailed in
Appendix I, from which we borrow two identities eq(\ref{id1a})
\eqa
\sum\limits_{m} \chi_{m_1m}(\vq_1)\rho_{mm_2}(\vq_2)=&\nonumber\\
<m_1|e^{-i({\vq_1\over c}+c\vq_2)\cdot\vR_{CF}^{c}}|m_2>
&e^{-{i\over2}\vQ_1\times\vQ_2}
\label{id1}\eea
and eq(\ref{id2a})
\eq
\sum\limits_{m} \chi_{m_1m}(\vq_1)\rho_{mm_2}(\vq_2)
=e^{-i\vQ_1\times\vQ_2}\sum_{m} \rho_{m_1m}(\vq_2)\chi_{mm_2}(\vq_1)
\label{id2}\ee
Let us now consider 
\eqa
&\sum\limits_{m_1m_2}^{}\chi_{m_1m_2}(\vq)\cH(m_1m_2;n_1n_2;\vq)=
(\e(n_1)-\e(n_2))&\chi_{n_1n_2}(\vq)\nonumber\\
&+\sum\limits_{m_1m_2}^{}(N_F(m_2)-N_F(m_1))\nonumber\\
&{v(q)\over2\pi(l^*)^2}e^{-q^2l^2/2} \chi_{m_1m_2}(\vq)\rho_{m_2m_1}(-\vq)\rho_{n_1n_2}(\vq)\nonumber\\
&-\sum\limits_{m_1m_2}^{}(N_F(m_2)-N_F(m_1))\ints v(s)e^{-{s^2l^2\over2}}\nonumber\\
&\rho_{n_1m_1}(\vs)\chi_{m_1m_2}(\vq)\rho_{m_2n_2}(-\vs)e^{i\vS\times\vQ}
\eea
Let us consider the direct and exchange terms separately. In the
direct term, one $m$ index can always be summed freely, while the
other is constrained by the Fermi occupation factor $N_F$. The sum
over the free $m$ gives, according to eq(\ref{id1})
\eqa
\sum\limits_{m_2}^{} &N_F(m_2)\rho_{n_1n_2}(\vq) <m_2|e^{-i({1\over c}-c)\vq\cdot\vR_{CF}^{c}}|m_2>\nonumber\\
-\sum\limits_{m_1}^{} &N_F(m_1)\rho_{n_1n_2}(\vq) <m_1|e^{-i({1\over c}-c)\vq\cdot\vR_{CF}^{c}}|m_1>
\eea
The two terms are immediately seen to cancel. Now let us turn to the
exchange terms, and consider the one that has the factor $N_F(m_2)$,
and a free sum over $m_1$. In this term, one can use eq(\ref{id2}) to exchange
the $\rho$ and $\chi$ matrix elements to obtain
\eqa
-\sum\limits_{m_2}^{}&N_F(m_2)\sum\limits_{m_1}\chi_{n_1m_1}(\vq)\times\nonumber\\
\ints v(s)&e^{-{s^2l^2\over2}} \rho_{m_1m_2}(\vs)\rho_{m_2n_2}(-\vs)
\eea
Notice that the phase factor $e^{i\vS\times\vQ}$ has been cancelled by
an opposite phase factor from eq(\ref{id2}). Now the angular $\vs$ integral
forces $m_1=n_2$ for a rotationally invariant potential, and the
result contains  the Fock energy of the state $n_2$
\eq
\e^{F}(n_2)\chi_{n_1n_2}(\vq)
\ee
Similarly, the other exchange term proportional to $N_F(m_1)$ ends up
giving $-\e^{F}(n_1)\chi_{n_1n_2}(\vq)$. Due to the peculiar nature of
the Hamiltonian the Hartree energy is a constant independent of the
CF-LL index, and the difference of the Fock energies is the same as
the difference of the full HF energies. Thus, the exchange
contributions cancel the diagonal term
$(\e(n_1)-\e(n_2))\chi_{n_1n_2}(\vq)$, and $\chi_{n_1n_2}(\vq)$ is
indeed an left eigenvector with zero eigenvalue for $\cH$, as claimed.

If we try to obtain another eigenvector with zero eigenvalue using
eqs(\ref{pm1},\ref{pm2}), we are foiled by the fact that
$\chi_{n_1n_2}(q_x)=\chi_{n_2n_1}(q_x)$, which gives us back the same
eigenvector. The matrix $\cH$ is {\it defective}, that is, it does not
have a complete set of eigenvectors. The null subspace of $\cH$ is
two-dimensional, one of the vectors being the eigenvector
$\Psi_{0}^{L}(m_1m_2;\vq)=\chi_{m_1m_2}(\vq)$. The other vector (call
it $\Psi_{0'}^{L}$) is such that $\Psi_{0'}^{L}\cH=\Psi_{0}^{L}$. In
other words, in this subspace, $\cH$ looks like
\eq
\left( \begin{array}{cc}
              1   & 0  \\
              1   & 0
           \end{array}\right)
\ee

Fortunately, this singular behavior of the matrix is not a problem in
computing physical correlators, because the density does not couple to
the null subspace, as will be seen explicitly in the small $q$ limit
in the next section. To compute physical correlators, one need only
project out this two-dimensional null subspace, and then the matrix
inverse $(\o-\cH)^{-1}$ looks exactly as in eq(\ref{specrep}), except
that the sum is only over physical eigenvalues.

\section{Small $q$ analysis of $\cH$}
\label{sec5}

The goal of this section is to compute the small-$q$ behavior of the
structure factor $S(q)$, which is the equal-time density-density
correlator in the ground state. To this end, we will carry out an
analysis of $\cH$ for small $q$, and show that one needs to keep at
most a $6\times6$ matrix to get the structure factor correct to order
$q^4$. One of the  results of this section will be that the
coefficient of this $q^4$ term diverges as $p\to\infty$, showing that
the limits $q\to 0$ and $p\to\infty$ do not commute. 

Noting that the density can be expressed as
\eq
\rho_e(\vq)=\sum\limits_{m_1m_2}e^{-q^2l^2/2} \rho_{m_1m_2}(\vq) O_{m_1m_2}(\vq)
\ee
we can write the density-density correlator
$S(\vq,\o)=<\rho_e(\vq,\o)\rho_e(-\vq,-\o)>$ as
\eqa
&S(\vq,\o)={L^2\over2\pi(l^*)^2}e^{-q^2l^2/2}\nonumber\\
&\sum\limits_{\a,m_1m_2} \rho_{m_1m_2}(\vq)\Psi_{\a}^R(\mu;\vq){1\over\o-E_{\a}+i\e sgn(E_{\a})}\nonumber\\
&\Psi_{\a}^L(n_1n_2:\vq)(N_F(n_2)-N_F(n_1))\rho_{n_2n_1}(-\vq)
\eea
where the sum is over physical states only, and the $i\e$ factors
appropriate to the time-ordered Green's function have been
inserted. Now one can go back to the time-domain Green's function, and
take the limit of equal time to write in a transparent notation
\eqa
S(q)=&{L^2\over2\pi(l^*)^2 N}e^{-q^2l^2/2}\nonumber\\
&\sum\limits_{\a,E_{\a}>0} <\rho(\vq)|\Psi_{\a}^R(\vq)><\Psi_{\a}^{L}(\vq)|\rho(\vq)>
\eea
where $N$ is the number of particles, and we have defined
\eqa
&(<\rho(\vq)|)_{m_1m_2}=&\rho_{m_1m_2}(\vq) \\
&(|\rho(\vq)>)_{m_1m_2}=&(N_F(m_2)-N_F(m_1))\rho_{m_2m_1}(-\vq)
\eea
Recalling the relation between the right and left eigenvectors
eq(\ref{psiright}), and noting that $N=pL^2/2\pi(l^*)^2$ for
$\nu={p\over2p+1}$, we finally obtain
\eq
S(q)={1\over
p}e^{-q^2l^2/2}\sum\limits_{\a,E_{\a}>0}|<\rho(\vq)|\Psi_{\a}^{R}(\vq)>|^2
\label{sofq}\ee

Our task in this section will be to compute all terms in this sum that
contribute to order $(ql)^4$. It turns out that the case of a single
CF-LL filled is nongeneric, while any $p>1$ ($\nu=p/2p+1$) is
generic. We address the simpler $p=1$ case first.

\subsection{The Structure factor for the Laughlin fractions}
\label{sec5.1}

From the structure of eq(\ref{sofq}) it is clear that only
magnetoexciton operators with $N_F(m_1)\ne N_F(m_2)$ need to be kept
in computing the density correlator. For $\nu=\third$ this corresponds
to $O_{m0}$ (positive energy) and $O_{0m}$ (negative energy). We will
assume that the eigenvectors $\Psi_{\a}$ are analytic at small $q$,
which turns out to be correct in the incompressible
states. Furthermore, we will need the matrix elements $\rho_{m0}$
which are
\eq
\rho_{m0}(\vq)=\sqrt{1\over m!} e^{im(\t_q-\pi/2)} \big({cQ\over\sqrt{2}}\big)^m e^{-c^2Q^2/4}
\ee
and $\rho_{0m}(\vq)=(\rho_{m0}(-\vq))^*$. It is clear that
$\rho_{10}\approx q$, and $\rho_{20}\approx q^2$, and the other matrix
elements go with higher powers of $q$. Therefore one needs to keep
only the components $10$, $20$, $01$, and $02$ of all the eigenvectors
$\Psi_{\a}$ to find $S(q)$ to order $q^4$.

In order to compute these components of the eigenvectors, we need to
analyse the small-$q$ structure of $\cH$ in greater detail. In the
following, we will set $\t_q=\pi/2$ for simplicity. This choice makes
$\cH$ real.

First consider the diagonal matrix elements.
\eqa
\cH(m0;m0;q)=&\e(m)-\e(0)+{v(q)\over2\pi(l^*)^2 m!}e^{-Q^2/2}\big({c^2Q^2\over2}\big)^m\nonumber\\
&-\int {sds\over(2\pi)} v(s)e^{-S^2/2}L_m({c^2S^2\over2})J_0(QS)
\eea
Let us recall that the differences of energies $\e(m)-\e(0)$ are
given in terms of the potential (eq(\ref{hfenergy}))
\eqa
\e(m)-\e(0)=&\ints v(s)e^{-s^2l^2/2} (|\rho_{00}(\vs)|^2-|\rho_{m0}(\vs)|^2)\nonumber\\
&=\int {sds\over(2\pi)} v(s)e^{-S^2/2} \big(1-{1\over m!}\big({c^2S^2\over2}\big)^m\big)
\eea

Consider $\cH(10;10,q)$. For small $q$ the direct term contributes
only in order $q$ (for a long-range potential, $v(q\to0)\simeq 1/q$), or
order $q^2$ (for a short-range potential, $v(q\to0)\simeq\ const$), while
the energy difference and the exchange term contribute to order
$q^0$. Setting $J_0\approx 1$ to get the order 1 contribution, we find
that it {\it vanishes for an arbitrary potential}! The other diagonal
matrix elements $m0$ for $m\ne1$ show no such behavior. This means
that the naive magnetoexciton $O_{10}$ has zero energy as
$q\to0$. This signals us that it must be related to the
constraint. The detailed manner in which it is related will be seen
shortly.

 As an aside we note that when one tries to construct
CF-magnetoexciton wavefunctions in the standard manner by taking a IQH
magnetoexciton, multiplying by the Laughlin-Jastrow factor, and
projecting, one finds\cite{jain-pvt} that the IQH $10$ magnetoexciton
projects to zero as $q\to0$. This seems to be directly related to what
we find here, but since it is difficult to go back to electronic
wavefunctions from the Hamiltonian theory, it is not clear how to
extend this relationship further.

Let us now consider the off-diagonal elements of $\cH$. There are two
different types, those confined to the positive energy subspace, and
those that go between the positive and negative energy subspaces. 
\eqa
\cH(m0:n0:q)={v(q)\over2\pi(l^*)^2 \sqrt{m!n!}}&e^{-Q^2/2}\big({cQ\over\sqrt{2}}\big)^{m+n}\nonumber\\
-(-1)^{n_<(m,n)+m}(sgn(n-m))^{n-m}&\int {sds\over(2\pi)} v(s)e^{-S^2/2}\nonumber\\
\times \sqrt{{n_<(n,m)!\over
 n_>(n,m)!}}\big({cS\over\sqrt{2}}\big)^{|n-m|}
L_{n_<}^{|n-m|}&\big({c^2S^2\over2}\big)
 J_{|n-m|}(QS)
\eea
The most important property to note is that the off-diagonal matrix
element vanishes as $q^{|n-m|}$ as $q\to0$ (this behavior comes from
the exchange term). Secondly, by exchanging $m$ and $n$, it can be
seen that $\cH_{++}$ is a symmetric matrix. Now for the second kind of
off-diagonal element
\eqa
\cH(m0;0n;q)=&(-1)^n{v(q)\over2\pi(l^*)^2 \sqrt{m!n!}}e^{-Q^2/2}\big({cQ\over\sqrt{2}}\big)^{m+n}\nonumber\\
&-(-1)^n\int {sds\over(2\pi)} v(s)e^{-S^2/2}{J_{m+n}(QS)\over\sqrt{m!n!}}
\eea
Once again, this vanishes as $q^{m+n-1}$ (long-range potential) or
$q^{m+n}$ (short-range potential). 

Thus, as $q\to0$, the naive magnetoexcitons $O_{m0}$, $O_{0m}$ become
decoupled for $m>1$, and therefore become the true eigenstates of
$\cH$. $O_{10}$ and $O_{01}$ have a diagonal energy of order $q$
(long-range potential), and are coupled by matrix elements of order
$q$, so they do not decouple. We will use these $q\to0$ naive
magnetoexcitons as labels for the true eigenstates, denoting the
positive energy ones by $\Psi_{m}$ and the negative energy ones by
$\Psi_{-m}$ (for $m>1$). In order to find the true eigenstates
$\Psi_{\a}$ at small $q$, one can do perturbation theory with $q$ as
the small parameter. 

To see which of the true eigenstates we need to keep, consider the
components of $\Psi_{3}$. We have already seen that we need
to restrict attention to the components $10$, $20$, $01$, and $02$,
due to the powers of $q$ already present in $\rho$. Due to the small
$q$ structure of $\cH$ it can be seen that
\eq
\Psi_{3}^{R}\approx|30>+\a q|20> +\b q |40> +\gamma q^2 |10> + \cdots
\ee
where $|m0>$ represent the naive magnetoexciton states. Therefore the
overlap $<\rho(q)|\Psi_{3}^{R}>\approx q^3$, and it is unnecessary to
compute it if one wants $S(q)$ to order $q^4$. Similar reasoning holds
for $\Psi_{m}$ for any $m>2$ and $m<-2$. 

To summarize, we see that to obtain $S(q)$ to order $q^4$ it is
sufficient to compute $|<\rho(q)|\Psi_{2}>|^2$, since $\Psi_1$ will
belong to the null (or unphysical) subspace.
Also, the overlap needs to be computed only to an accuracy
of $q^2$. 

To make further progress, let us write down the matrix $\cH$
restricted to the lowest-energy four-dimensional subspace spanned by
$|20>$, $|10>$, $|01>$, and $|02>$, where all the diagonal matrix
elements have been kept to order $q^2$, and a long-range interaction
has been used.
\eq
\cH_4=\left(\begin{array}{cccc}
                E_{20}(k)& Bk&O(k^2)&O(k^3)\\
                Bk&Ak&Ak-Ck^2&O(k^2)\\
                O(k^2)&-Ak+Ck^2&-Ak&Bk\\
                O(k^3)&O(k^2)&Bk&-E_{20}(k)
             \end{array}\right)
\ee
We have defined $k=ql$. The various coefficients are (for the Coulomb
interaction with the energy scale $e^2/\e l$ removed, $v(q)=2\pi l/q$)
\eqa
A=&\third\nonumber\\
B=& -{1\over4}\sqrt{{\pi\over3}}  \nonumber\\
C=&{3\over8}\sqrt{{\pi\over6}}\nonumber\\
E_{20}=&\third \sqrt{{\pi\over6}}+O(q^2)
\eea

The central $2\times2$ submatrix consisting of the states $10$ and
$01$ can be seen to be very singular. Taken by itself, it possesses
two zero eigenvalues (up to order $q^2$) but only one eigenvector, and
represents the constraint in its small-$q$ form. The larger matrix
also possesses two zero eigenvectors.  This can be seen from the
determinant of $\cH_4$, which is
\eq
det|\cH_4|=2AE_{20} k^3(CE_{20}-B^2)+k^4(B^4-C^2E_{20}^2)
\ee
The key identity 
\eq
B^2=CE_{20}(0)
\label{bceid}\ee
 makes this determinant zero to order $k^4$ inclusive, which means
that there are two zero eigenvalues to order $k^2$. Eq(\ref{bceid})
seems accidental, and dependent on the potential used, but it can be
shown from the form of the low-order Laguerre polynomials that this
identity holds independent of the potential, and furthermore that
\eq
B=-{3\over 2\sqrt{2}} E_{20}(0)
\ee
for any potential.

As mentioned before, the true eigenvectors can be found by
straightforward perturbation theory. We first diagonalize the positive
energy subspace by a unitary transformation (the Hamiltonian is
hermitian in this subspace). The eigenvectors can be used to construct
the unitary transformation which needs to be correct only to order $k$
\eq
U=\left(\begin{array}{cc}
        1&-{Bk\over E_{20}}\\
       {Bk\over E_{20}}&1
       \end{array} \right)
\ee
The resulting eigenvalues are $E_{20}+{B^2k^2\over E_{20}}\equiv E$ and
$Ak-B^2k^2/E_{20}\equiv Ak-Ck^2$, and the matrix in the new basis
looks like
\eq
U^{\dagger}\cH_4 U=\left(\begin{array}{cccc}
           E&0&O(k^2)&O(k^3)\\
           0&Ak-Ck^2&Ak-Ck^2&O(k^2)\\
           O(k^2)&-Ak+Ck^2&-Ak+Ck^2&0\\
           O(k^3)&O(k^2)&0&-E
	   \end{array}\right)
\ee
It is clear that to order $q$ the nonzero positive eigenvector is
decoupled from the negative energy subspace. The central $2\times2$
matrix is degenerate, and has two zero eigenvalues, but only one
eigenvector. It represents the null subspace of $\cH$ in the new
basis. In order to compute $S(q)$ to order $q^4$ we need only the
physical positive energy eigenvector $\Psi_{2}$. It can be seen that
$\Psi_2$ is decoupled from the null subspace. To the requisite order
$\Psi_2$ is
\eq
\Psi_{2}^{L}=\left(\begin{array}{cccc}
             1&{Bk\over E_{20}(0)} & O(k^2)&O(k^3)
\end{array}\right)
\ee
It can be easily verified that $\Psi_{2}$ is orthogonal to the null
eigenvector, and this fact provides a shortcut to deriving the above
results.  Now taking the overlap with $|\rho>$ and squaring, one
easily finds
\eq
S(q)={(ql)^4\over8}+\cdots
\ee

This coefficient differs by a factor of 2 from that predicted from
Laughlin's wave function\cite{laugh} using the compressibility sum
rule\cite{GMP}, namely $S(q)=(ql)^4/4$. Therefore this calculation,
while unambiguously showing the signature of an incompressible state,
does not give the same result as the Laughlin state. This is a little
surprising, since one expects the CF-HF state of one CF-LL filled to
correspond to the Laughlin state. However, two points should be noted
here. Firstly, it is very difficult to make the correspondence between
the CF-wavefunction and the electronic one in the Hamiltonian
formalism. It may be that the electronic wavefunction corresponding to
one full CF-LL is incompressible, but not Laughlin's
wavefunction. Secondly, this is an approximate calculation, albeit one
that is conserving, and a more sophisticated conserving approximation
may well produce the same answer as Laughlin's wave function.

Finally, one may extend this result to all the Laughlin fractions
${1\over2s+1}$ by attaching $2s$ units of flux, with
$c^2=\sqrt{{2s\over2s+1}}$ to obtain
\eq
S(q)={1\over 8} (ql)^4 +\cdots
\ee
independent of $s$, which also differs from the result for the
coefficient obtained from Laughlin's wavefunction\cite{GMP}, namely
$s/4$.

It should be emphasized that the results are independent of the form
of the potential. The short-range potential produces a slightly
different form of the matrix $\cH_4$ but one still obtains the same
answer for the small $q$ behavior of $S(q)$. One expects this since
the structure factor is a property of the ground state, and should not
depend on the dynamics.

\subsection{The Structure Factor for the Principal Fractions $\nu=p/2p+1$, $p>1$}
\label{sec5.2}

Let us now turn to the more generic case of $p>1$. In this case we
have $p$ CF-LLs filled. We have already seen that only naive
magnetoexcitons with $N_F(m_2)\ne N_F(m_1)$ contribute to the density
correlator, so we will only consider those below. Once again, we find
that the diagonal matrix element $\cH(p,p-1;p,p-1;\vq)$, which is
\eqa
&\e(p)-\e(p-1)+{v(q)\over 2\pi(l^*)^2p}e^{-Q^2/2} \big({cQ\over\sqrt{2}}\big)^2\nonumber\\
&-\int {sds\over 2\pi} v(s)e^{-S^2/2} L_{p-1}\big({c^2S^2\over2}\big)L_{p}\big({c^2S^2\over2}\big)J_0(QS)
\eea
vanishes as $q\to0$. Recall that the energy difference is (eq(\ref{hfenergy}))
\eqa
&\e(p)-\e(p-1)=\intq v(q)e^{-Q^2/2}\sum\limits_{k=0}^{p-1}\nonumber\\
& \big({k!\over(p-1)!}\big({c^2Q^2\over2}\big)^{p-1-k} \big(L_{k}^{p-1-k}\big)^2-{k!\over p!}\big({c^2Q^2\over2}\big)^{p-k} \big(L_{k}^{p-k}\big)^2\big)
\eea
The $p,p-1$ diagonal matrix element vanishes as $q\to0$ for an {\it
arbitrary} potential by virtue of a Laguerre polynomial identity
\eqa
\sum\limits_{k=0}^{p-1}{k!\over(p-1)!}x^{p-1-k} \big(L_{k}^{p-1-k}(x)\big)^2&-{k!\over p!}x^{p-k} \big(L_{k}^{p-k}(x)\big)^2\nonumber\\
&=L_{p-1}(x)L_p(x)
\eea
This identity has been verified to high order $p$, but a general proof
is not known to the author.

Once again, this vanishing energy is related to the
constraint. Proceeding to the modes that are connected to this one by
order $q$ matrix elements, we find that there are two, $O_{p+1,p-1}$
and $O_{p,p-2}$. This is the primary difference between the case of
generic $p$ and $p=1$, since in the latter case the second operator
was missing. Now one proceeds as before, by first looking at  the
positive energy subspace 
\eq
\cH_{++,3}=\left(\begin{array}{ccc}
                 E_1(q)&V_{12}(q)&B_1k\\
                 V_{12}(q)&E_2(q)&B_2k\\
                 B_1k&b_2k&E_0(q)
                \end{array}\right)
\ee
where 
\eqa
E_0=&\cH(p,p-1;p,p-1;q)\\
E_1=&\cH(p+1,p-1;p+1,p-1;q)\\
E_2=&\cH(p,p-2;p,p-2;q)\\
V_{12}=&\cH(p+1,p-1;p,p-2;q)\\
\cH(p,p-1&;p+1,p-1;q)=B_1k+\cdots\\
\cH(p,p-1&;p,p-2;q)=B_2k+\cdots
\eea
It should be noted that $V_{12}$ does not vanish as $q\to0$.  Next one
carries out a unitary transformation that makes the upper left
$2\times2$ submatrix diagonal. This unitary transformation is 
\eq
U_1={1\over D}\left(\begin{array} {ccc}
                    V_{12}& -(E_+-E_1)& 0\\
                    E_+-E_1&V_{12}&0\\
                    0  &0  &  D
                    \end{array}\right)
\ee
where 
\eqa
E_{\pm}=&{1\over2}(E_1+E_2\pm\sqrt{(E_1-E_2)^2+V_{12}^2})\\
D=&\sqrt{(E_+-E_1)^2+V_{12}^2}
\eea
After this transformation the Hamiltonian in the $3\times3$ positive
energy subspace assumes the form 
\eq
\cH_{++,3}=
\left(\begin{array}{ccc}
                  E_+&0&k{\tilde{B}}_1\\
                 0&E_-&k{\tilde{B}}_2\\
k{\tilde{B}}_1&k{\tilde{B}}_2&0
            \end{array}\right)
\ee
where 
\eqa
{\tilde{B}}_1=&{V_{12}B_1+(E_+-E_1)B_2\over D}\\
{\tilde{B}}_2=&{V_{12}B_2-(E_+-E_1)B_1\over D}
\eea
Let us call the basis vectors here $\Psi_+$, $\Psi_-$, and $\Psi_0$.
Now it is easy to construct new basis vectors $\widetilde{\Psi}_+$,
$\widetilde{\Psi}_-$ and $\widetilde{\Psi}_0$ in the positive energy subspace
that make it completely diagonal to order $q^2$ by doing first order
perturbation theory in the off-diagonal elements. Once again it turns
out that with this transformation, the positive energy subspace is
decoupled from the negative energy and null subspaces to order
$q^2$. The structure factor is then computed from the overlaps of the
density with the physical positive energy eigenvectors.
\eqa
&S(q)={1\over p}\big(|<\rho(q)|\widetilde{\Psi}_+(q)>|^2+|<\rho(q)|\widetilde{\Psi}_-(q)>|^2\big)=\nonumber\\
&(ql)^4p\big[\big({1\over\sqrt{p+1}}+{E_2B_1-V_{12}B_2\over E_1E_2-V_{12}^2}\big)^2
+\big({1\over\sqrt{p-1}}+{E_1B_2-V_{12}B_1\over E_1E_2-V_{12}^2}\big)^2\big]
\label{sfp2}\eea

All the above algebraic steps are heavily dependent on the potential
used, but nevertheless one expects that the final result should be
potential independent. Once again, a formal proof is lacking, but this
potential-independence has been verified for the Coulomb potential and
the short range Gaussian potential for $2/5$ and $3/7$. By some
mysterious (to the author) properties of the Laguerre polynomials and
their integrals, all the dependence on the functional form of the
potential drops out.

Let us illustrate this with the Gaussian potential $v(q)=2\pi(l^*)^2
e^{-\a Q^2/2}$, which is very convenient for explicitly carrying out
integrals, but which (for large $\a$) also possesses a small parameter
$\b=1/(1+\a)$ which will be exploited later. For $p=2$ the relevant
quantities (without any approximation) are
\eqa
E_1=&{12\over5}\b^2-{32\over5}\b^3+{192\over25}\b^4-{2048\over625}\b^5\\
E_2=&{4\over5}\b^2+{32\over25}\b^3-{192\over125}\b^4\\
V_{12}=&-{1\over\sqrt{3}}\big({12\over5}\b^2-{96\over25}\b^3+{192\over125}\b^4\big)\\
B_1=&-{\b^2\over\sqrt{3}}\big(3-{48\over5}\b+{336\over25}\b^2-{768\over125}\b^3\big)\\
B_2=&\b^2\big(1-{16\over5}\b+{48\over25}\b^2\big)
\eea
The expression appearing in the denominator of eq(\ref{sfp2}) is
\eqa
E_1E_2-V_{12}^2=&\b^4\big({2^9\over5^3}\b-{2^{13}\over5^4}\b^2+{2^{16}\over5^5}\b^3\nonumber\\
&-{2^{18}\over5^6}\b^4+{3\times2^{17}\over5^7}\b^5\big)
\eea
It can easily be verified that both $E_1B_2-V_{12}B_1$ and
$E_2B_1-V_{12}B_2$ are proportional to the above expression, giving 
ratios independent of $\b$. The structure factor depends on these
ratios only, and is thus independent of $\b$.

For $p>3$ it becomes very difficult to compute $E_i$, $B_i$, and
$V_{12}$ exactly, even for the Gaussian potential. However, if one
{\it assumes} the potential independence of $S(q)$, then one can use
the following strategy: Since $S(q)$ does not depend on $\b$, it can
be calculated in the limit of small $\b$. In this limit, one need only
compute terms of lowest nontrivial order in $\b$ in $E_i$, $B_i$, and
$V_{12}$. Keeping one more order as a consistency check one finds for generic $p$
\eqa
E_1=&\b^2c^2(p+1)-2\b^3c^4(p^2+p-1)+\cdots\\
E_2=&\b^2c^2(p-1)-\b^3c^4(2p^2-6p+2)+\cdots\\
V_{12}=&-\sqrt{p^2-1}(\b^2c^2-2\b^3c^4(p-1)+\cdots)\\
B_1=&\sqrt{p+1}\b^2(1-\b c^2(3p-2)+\cdots)\\
B_2=&\sqrt{p-1}\b^2(1-\b c^2(3p-2)\cdots)
\eea

Using the above expressions in eq(\ref{sfp2}) one finds the result quoted
in the introduction
\eq
S(q)={(ql)^4\over2}{p^4-3p^3+{5\over4}p^2+3p+{7\over4}\over p^2-1}
\label{sfp3}\ee
This expression, with its divergence as $p\to\infty$ or $\nu\to\half$
is consistent with the result\cite{read3} that for a problem
equivalent to the $\nu=\half$ problem, $S(q)\simeq q^3\log(q)$. As
$p\to\infty$, the radius of convergence of the power series expansion
of $S(q)$ must go to zero (else the structure factor would diverge for
a range of $q$ according to eq(\ref{sfp3})), and the $p\to\infty$ limit
does not commute with the $q\to0$ limit.

Finally, it is interesting to compare the conserving prediction for
the $q^4$ term of $S(q)$ with the naive calculation using the
preferred density in the HF state
\eq
S_{naive}(q)=<\Omega_{HF}|\rho_{p}(\vq)\rho_{p}(-\vq)|\Omega_{HF}>
\ee

This latter number turns out to be $p/8$. Read\cite{read3} wrote the
conserving density-density correlator as a sum of the naive correlator
of the preferred density and a term representing the coupling of this
preferred density to a gauge field. Making a similar decomposition
here we see that the gauge-field contribution increases with $p$, and
comes to dominate the small-$q$ structure factor as $p\to\infty$.

\section{Explicit results for $\third$ and $\twof$}
\label{sec6}

To motivate the results presented in this section let us consider the
relation of the all-$q$ Hamiltonian theory\cite{aftermath} to the
original electronic theory.  The all-$q$ theory is algebraically
consistent at all $q$ and has the same algebra for the electronic
density as the electronic theory. It is also consistent with the
small-$q$ theory that was developed earlier by canonically
transforming the electronic Hamiltonian. If one could find a sequence
of canonical transformations that exactly mapped the electronic
Hamiltonian into this one, then one could be confident that the two
problems are identical. In the absence of such a transformation, all
we can really be sure of is that the all-$q$ theory has the same
small-$q$ behavior as the electronic problem. It is therefore of
interest to compute the magnetoexciton dispersions for $q$ not very
small to see if it agrees with the numerical results for the
electronic problem. In this section we will present results for the
spin-polarized magnetoexcitons for $\third$ and $\twof$ and show that
the minima in the lowest energy magnetoexciton dispersion occur at the
same values of $ql$ as those found for the electronic problem.

As shown in the previous section, for very small $q$, the naive
magnetoexcitons decouple and become the true eigenstates of $\cH$. As
$q$ increases they become increasingly coupled, and both level
repulsion (between positive energy states) and level attraction
(between positive and negative energy states) manifest themselves,
giving rise to complicated magnetoexciton dispersions. The matrix
$\cH$ is infinite-dimensional, but in any numerical calculation, only
a finite matrix can be diagonalized. We saw that the lowest nontrivial
result for $S(q)$ could be obtained by keeping at most a $6\times6$
matrix. As $q$ increases more and more CF-LLs have to be kept to
obtain accurate results. The accuracy of the results were checked by
two different methods. Firstly, the number of CF-LLs kept was
increased until the energy of the magnetoexciton was stable. Secondly,
we know that at every $q$ there should be two zero eigenvalues
corresponding to the unphysical sector. For a finite number of CF-LLs
kept, and $q$ not too small, these would-be zero eigenvalues are not
exactly zero, but much smaller than any other eigenvalues. The number
of CF-LLs kept in the calculation was increased until these null
eigenvalues were at least 4 orders of magnitude smaller than the
smallest physical eigenvalue.

\begin{figure}
\narrowtext
\epsfxsize=2.4in\epsfysize=2.4in
\hskip 0.3in\epsfbox{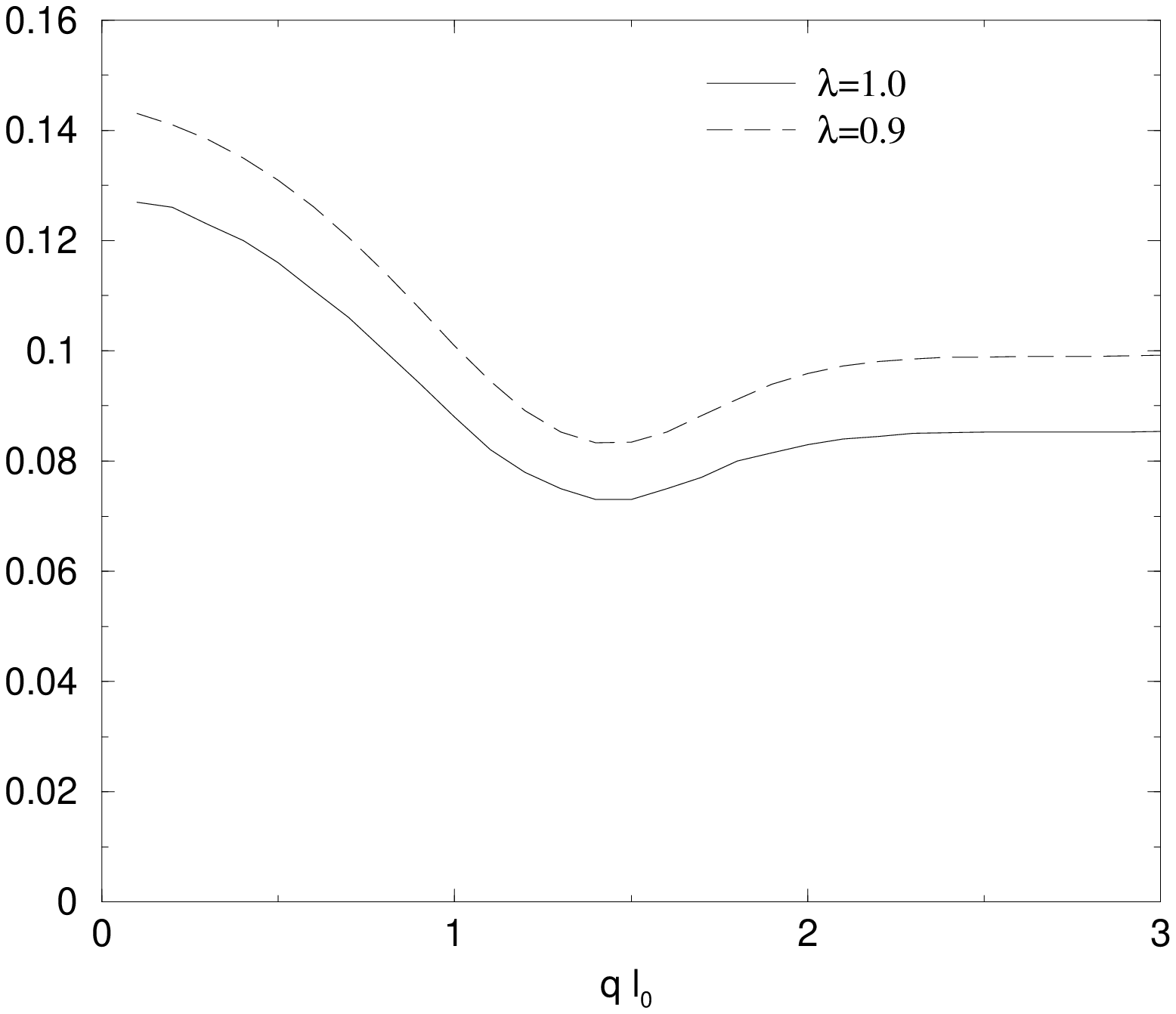}
\vskip 0.15in
\caption{Spin-polarized magnetoexciton dispersions for $\nu=\third$ 
at $\lam=0.9l$ and $\lam=l$.  All energies are in units of
$e^2/\varepsilon l$. 
\label{fig1}}
\end{figure}

The two figures are for the long-range potential $v(q)={2\pi e^2\over
\e q}e^{-q^2\lambda^2/2}$ for various values of $\lambda$. It is worth
noting that though the depths of the minima are $\lambda$-dependent,
and different from what are found in numerical diagonalizations or
from CF-wavefunctions, the positions are
correct\cite{jain-cf-review,hal-rez,exact-ex,jain-ex}. This indicates that the
all-$q$ theory does capture some of the important physics of the
electronic problem even at fairly large $q$. Furthermore, the results
are calculated in the lowest level conserving approximation. A more
sophisticated conserving approximation incorporating the screening of
the potential may produce better results at large $q$.

\begin{figure}
\narrowtext
\epsfxsize=2.4in\epsfysize=2.4in
\hskip 0.3in\epsfbox{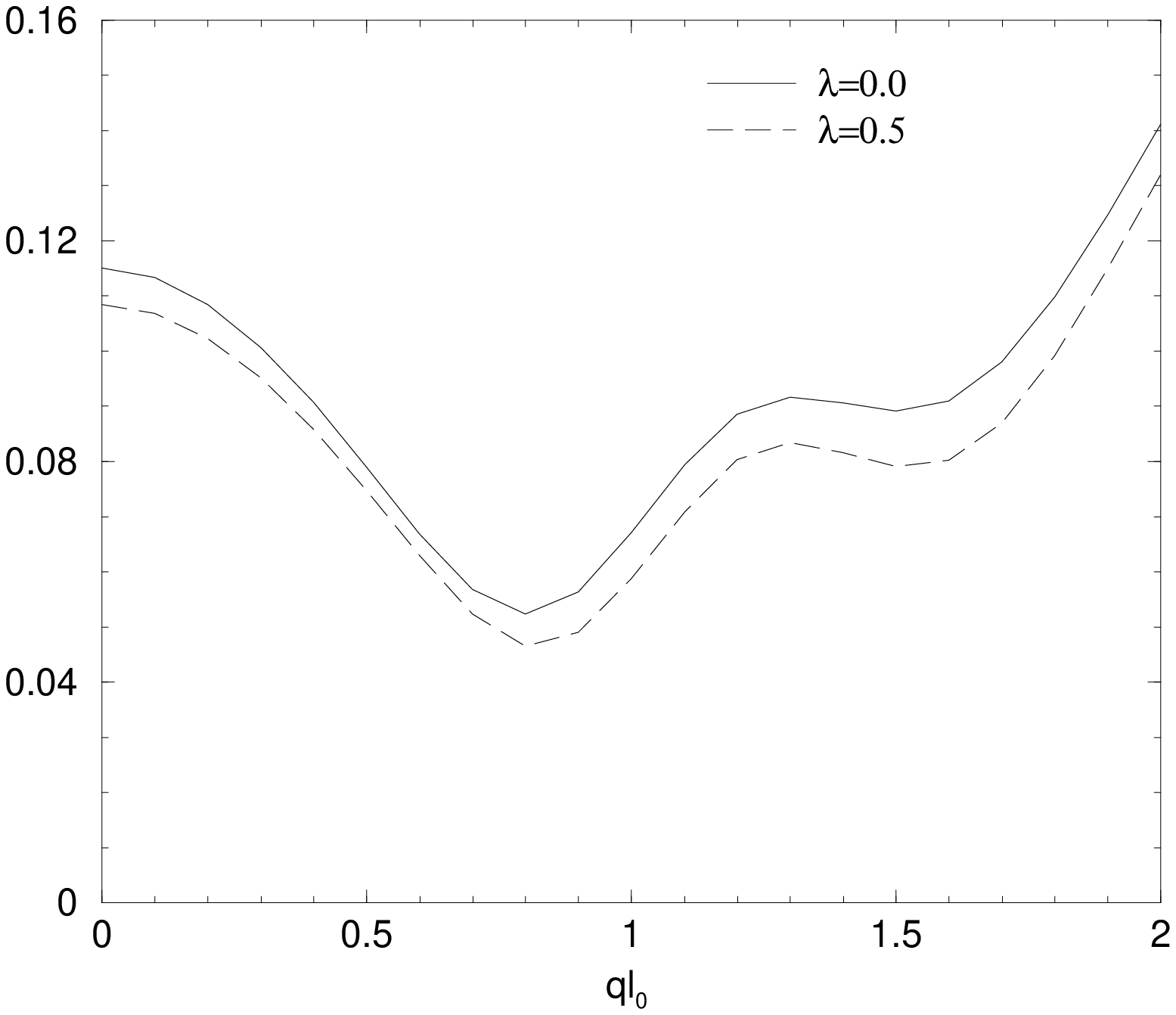}
\vskip 0.15in
\caption{Spin-polarized magnetoexciton dispersions for $\nu=\twof$ at
 $\lam=0$ and $\lam=0.5l$. All energies are in units of
$e^2/\varepsilon l$. 
\label{fig2}}
\end{figure}

\section{Extention of the formalism to nonuniform density}
\label{sec7}

In this section we will see how the fractional quasiparticle charge
emerges from the formalism, provided flux attachment is carried out
for the inhomogeneous problem. The all-$q$ Hamiltonian
theory\cite{aftermath} is based on the expectation that the ground
state is uniform. In the presence of a localized quasiparticle this is
clearly not a good assumption. More generally, it would be useful to
have a conserving formulation which was applicable to the case of a
nonuniform ground state, since this would enable a treatment of
edges\cite{edge-intro} or disorder.

To construct such a formalism, let us go back to the defining
equations for the guiding center coordinates of the electron and the
pseudovortex eq(\ref{redef},\ref{rvdef}). Let us assume that the
ground state density is $n(\vrr)=\bar{n}+\delta n(\vrr)$. According to
the principles of flux attachment, the CFs should see an additional
effective gauge field corresponding to this nonuniform density which
satisfies
\eq
\nabla\times\delta\vec{a}=4\pi\delta n(\vrr)
\ee
This additional gauge field should be present in the CF-velocity
operator $\vPi$, leading to the commutation relation
\eq
\left[\Pi_{i\a},\Pi_{i\b}\right]={i\e_{\a\b}\over l^2}(1-c^2-4\pi l^2\delta n(\vrr_i))
\label{pipi}\ee

With this change, the quasiparticle charge can be computed by a very
simple argument, identical in spirit to the counting argument given by
Jain\cite{jain-count}. Consider a region with large area $A$ which
contains one extra localized quasiparticle (in addition to $p$ filled
CF-LLs) in its interior. The total charge in this region (after
subtracting the background) is
\eq
Q=-e\delta n(q\to0)=-e(1+\delta n_R(q\to0))
\ee
where the 1 counts the extra CF that has been introduced, and $\delta
n_R$ is the response of the filled CF-LLs to the introduction of the
extra CF. Before the extra CF was introduced, all the states in the
CF-LLs $0,1,\cdots,p-1$ were occupied. If one assumes that the final
state is adiabatically connected to the initial one, then the same
states continue to be occupied. However, the total effective flux
through the area $A$ has decreased in magnitude by $4\pi\delta
n(q\to0)$. This means that each CF-LL has $2\delta n(q\to0)$ fewer
states in the area A than it did before. In essence, the states in the
regions of smaller effective field have a larger cyclotron radius, so
they expand and push out the other states. This means the total charge
in the area A is
\eq
Q=-e\delta n(q\to0)=-e(1-2p\delta n(q\to0))
\ee
which can be seen to give 
\eq
Q={-e\over 2p+1}
\ee

The inclusion of the varying gauge field in $\vPi$ is essential for
the success of this argument. Unfortunately, this modification of
$\vPi$ has an undesirable effect: Using the modified $\vPi$ in
the expressions for the guiding center coordinates $\vR_e$
(eq(\ref{redef})) and $\vR_v$ (eq(\ref{rvdef})) spoils their
commutation relations
(eqs(\ref{crrere},\ref{crrerv},\ref{crrvrv})). This is a serious
problem, since this means that the density operator constructed from
the new $\vR_e$ does not obey the magnetic translation algebra, and
thus cannot be identified with the physical electronic density. The
density also fails to commute with the constraint, making a conserving
approximation infeasible.

However, there is a way to modify $\vR_e$ and $\vR_v$ so that the
algebraic consistency of the theory is restored. Of the three
commutation relations (eqs(\ref{crrere},\ref{crrerv},\ref{crrvrv}))
only the first two are essential to having a consistent
theory. Satisfying eq(\ref{crrere}) would mean that the density
operator constructed from $\vR_e$ could be identified with the
electronic density, while satisfying eq(\ref{crrerv}) would mean that
this density commutes with the constraint. If eq(\ref{crrvrv}) is not
satisfied it means that the constraint algebra does not close. New
constraints would be generated by commuting old ones. This is not a
problem for a conserving approximation, since by the Jacobi identity
the electronic density operator (and thus the Hamiltonian constructed
from it) would commute with the new constraints as well.

It turns out that operators $\vR_e$, $\vR_v$ which satisfy the
conditions of the above paragraph to first order in ${\delta n\over
n}$ can be constructed. They are
\eqa
R_{e\a}=&r_{\a}+{l^2\over1+c}X_{\a} -
{2\pi l^4\over c(1+c)^2}\int\limits_{0}^{1} dt (1-t)\times\nonumber\\
&\big(X_{\a}\delta
n\big(\vrr-{tl^2\over c(1+c)}\vX\big)+\delta n\big(\vrr-{tl^2\over c(1+c)}\vX\big) X_{\a}\big)\label{remod}\\
R_{v\a}=&r_{\a}-{l^2\over c(1+c)}X_{\a} +
{2\pi l^4\over c^3(1+c)^2}\int\limits_{0}^{1} dt (c+t)\times\nonumber\\
&\big(X_{\a}\delta
n\big(\vrr-{tl^2\over c(1+c)}\vX\big)+\delta n\big(\vrr-{tl^2\over c(1+c)}\vX\big) X_{\a}\big)\label{rvmod}
\eea
where
\eqa
X_{\a}=&\e_{\a\b}\Pi_{\b}&\\
\delta n(\vrr-z\vX)=&\intq& \delta n(\vq) e^{i\vq\cdot(\vrr-z\vX)}
\eea
and $\delta n(\vq)$ is the Fourier transform of the ground state
expectation value of the electronic density operator, and is a
commuting number. The $\vPi$ operators appearing in the above formulas
contain the spatially varying gauge field $\delta\va$. The
preservation of the commutation relations
eqs(\ref{crrere},\ref{crrerv}) to first order in $\delta n/n$ can be
verified in a straightforward manner. This is therefore a good
representation of the electronic problem for small deviations of
density from a liquid state. Presumably eqs(\ref{remod},\ref{rvmod})
can be extended to all orders in $\delta n/n$ while still maintaining
the crucial commutation relations, eqs(\ref{crrere},\ref{crrerv}).

The modified density operator can in principle be used to calculate
the detailed charge profile of a quasiparticle. One can create an
extra CF in the $p^{th}$ CF-LL, and solve the CFHF equations with the
above expression for the density operator. There is a
self-consistent aspect to the formulation: The $\delta n(\vq)$ that
enters the density operator has to ultimately be the expectation value
of the same density operator in the ground state.

The modifications to $\vR_e$ and $\vR_v$ to order $\delta n/n$ (other
than the change in $\vPi$) are not relevant for the charge of an
isolated quasiparticle, but they are crucial if one wants to compute
bosonic response functions for a nonuniform state in a conserving
approximation. This issue will be of particular importance in the edge
state problem\cite{edge-intro}. There one knows that there are {\it
physical} gapless modes\cite{edge-intro}, and separating them from the
unphysical modes can only be reliably carried out within a
conserving approximation.

\section{Discussion and Conclusions}
\label{concl}

The fractional quantum Hall states\cite{fqhe-ex} have long had the
peculiar status of being understood in principle\cite{laugh,jain-cf}, and yet
being resistant to analytical calculation of their response
functions. In the past few years much understanding has been acheived
on how to compute these response functions. This understanding has
largely been based on the notion of field-theoretic flux attachment
via a Chern-Simons
transformation\cite{gcs,gmcs,zhk,read1,lopez,kalmeyer,hlr}. Outstanding
issues in the initial CS formulations\cite{ssh} were partially solved
by Shankar and the author by adding variables and carrying out a
sequence of canonical transformations to decouple the high and low
energy degrees of freedom\cite{us1,us2}. However, the resulting
theory, though it had many desirable features at small $q$, was
unsatisfactory when viewed as a theory of the Lowest Landau
level. Shankar posited a formulation\cite{aftermath} which was
algebraically consistent at all $q$ as a LLL theory, and had all the
desirable features of the small-$q$ formulation.

In the past, analysis of this all-$q$ formalism\cite{aftermath} was
limited to the HF approximation in which one used the preferred
density $\rho_p=\rho_e-c^2\rho_v$. The preferred density makes
explicit many nonperturbative properties of the true density operator,
such as the fractional quasiparticle charge and the order $q^2$ matrix
elements out of the ground state for incompressible states. The price
paid is that the constraints, once they are used to construct
$\rho_p$, are ignored, and one is not sure how robust the results are
under a proper implementation of the constraints.

The most important objective acheived in this paper is that the
nonperturbative features that were introduced by hand into the
preferred density\cite{us2} are seen to emerge very naturally from the
theory when constraints are consistently imposed and flux attachment
is carried out  for states of nonuniform density.

The conserving approximation used is identical in spirit to the one
used by Read\cite{read3} to obtain response functions for the $\nu=1$
boson problem based on the formalism of Pasquier and
Haldane\cite{pasquier}. The case considered here is in some ways more
complicated in that it involves Landau level states instead of plane
wave states, but also in other ways simpler, in that
finite-dimensional matrix equations have to be solved to obtain the
response rather than integral equations\cite{read3}.

The order $q^2$ matrix elements of the charge density operator emerge
when unphysical states are discarded and only physical states are
considered. The crux of the matter is that for $\nu={p\over2p+1}$ only
the $p-1$ to $p$ transition can produce an order $q$ matrix element in
violation of Kohn's theorem, and as $q\to0$ this magnetoexciton state
becomes identical to the constraint to order $q$, and thus unphysical.

Once the formalism is extended to include variations of density in the
ground state, the fractional quasiparticle
charge\cite{laugh,frac-charge,jain-count} also emerges from the theory.

Finally, a look at the spin-polarized magnetoexciton dispersions for
$q$ not too small shows that though one does not yet have a rigorous
connection between the all-$q$ theory and the original electronic
theory, the all-$q$ conserving approximation manages to capture much
of the important physics, even at large $q$.

In summary, it appears that the all-$q$ Hamiltonian formulation of the
fractional quantum Hall problem coupled to a conserving approximation
is well-suited to computing bosonic response functions in the
spin-polarized case. An analysis of the edge and disordered problems
also seems possible in this approach using the extention to small
density variations presented in Section\ \ref{sec7}, and will be
pursued in  future work.

\section{Acknowledgements}
I would like to thank J.K.Jain and R.Shankar for many helpful
conversations, and the NSF for partial support (DMR-0071611).

\section{Appendix I}
\label{app1}

Most of the material in this appendix appears in
ref.[\cite{full-disc}], but is presented here to make the discussion
self-contained. In order to compute the sums over CF-LL indices of
products of $\rho_e$ or $\rho_v$ matrix elements, it turns out to be
convenient to express all the coordinates in terms of the guiding
center and cyclotron coordinates of the CF, which are
\eqa
\vR_{CF}^{g}=&\vrr-{l^2\over(1-c^2)}\hz\times\vPi\\
\vR_{CF}^{c}=&{l^2\over(1-c^2)}\hz\times\vPi
\eea
which have the commutation relations
\eqa
\left[R_{CF,\a}^{c},R_{CF,\b}^{c}\right]=&i{l^2\over(1-c^2)}\e_{\a\b}\label{ccrc}\\
\left[R_{CF,\a}^{g},R_{CF,\b}^{g}\right]=&-i{l^2\over(1-c^2)}\e_{\a\b}\\
\left[R_{CF,\a}^{c},R_{CF,\b}^{g}\right]=&0
\eea
We can now express the electron and pseudovortex guiding center
coordinates in terms of the CF operators $\vR_{CF}^{c}$ and
$\vR_{CF}^{g}$ as
\eqa
\vR_{e}=&\vR_{CF}^{g}+c\vR_{CF}^{c}\\
\vR_{v}=&\vR_{CF}^{g}+{1\over c}\vR_{CF}^{c}
\eea
The CF guiding center and cyclotron coordinates, in turn, can be
expressed in terms of two sets of canonical, commuting creation and
annihilation operators
\eqa
R_{CF,x}^{c}+iR_{CF,y}^{c}=&l\sqrt{{2\over1-c^2}}a_c\\
R_{CF,x}^{c}-iR_{CF,y}^{c}=&l\sqrt{{2\over1-c^2}}a_c^{\dagger}\\
R_{CF,x}^{g}+iR_{CF,y}^{g}=&l\sqrt{{2\over1-c^2}}a_g^{\dagger}\\
R_{CF,x}^{g}-iR_{CF,y}^{g}=&l\sqrt{{2\over1-c^2}}a_g
\eea
where
\eqa
\left[a_c,a_c^{\dagger}\right]=&\left[a_g,a_g^{\dagger}\right]=1\\
\left[a_c,a_g^{\dagger}\right]=&\left[a_g,a_c^{\dagger}\right]=0
\eea
We can abstractly set up the states of the CF-cyclotron creation and
annihilation operators
\eq
|n>={(a_c^{\dagger})^n\over\sqrt{n!}}|0>
\ee
In terms of these states the matrix elements of eq(\ref{rhoe12}) and
eq(\ref{rhov12}) are simply expressed as
\eqa
\rho_{m_1m_2}(\vq)=&<m_1|e^{-ic\vq\cdot\vR_{CF}^{c}}|m_2>\\
\chi_{m_1m_2}(\vq)=&<m_1|e^{-{i\over c}\vq\cdot\vR_{CF}^{c}}|m_2>
\eea
In other words, only the cyclotron part of the electron and
pseudovortex density enters the matrix elements defined above. An
important result we will need  is the following sum
\eq
\sum_{m} \chi_{m_1m}(\vq_1)\rho_{mm_2}(\vq_2)
\ee
Using the completeness of the cyclotronic states $|m>$ we can recast this sum as
\eq
<m_1|e^{-{i\over c}\vq_1\cdot\vR_{CF}^{c}}e^{-ic\vq_2\cdot\vR_{CF}^{c}}|m_2>
\ee
We can then use the commutation relations of eq(\ref{ccrc}) to combine
the exponentials and obtain
\eqa
\sum_{m} \chi_{m_1m}(\vq_1)\rho_{mm_2}(\vq_2)=&
e^{-{i\over2}\vQ_1\times\vQ_2}\times\nonumber\\
&<m_1|e^{-i({\vq_1\over c}+c\vq_2)\cdot\vR_{CF}^{c}}|m_2>
\label{id1a}\eea
Finally we 
separate the exponentials in the reverse order to get the second useful
identity
\eqa
&\sum_{m} \chi_{m_1m}(\vq_1)\rho_{mm_2}(\vq_2)\nonumber\\
&=e^{-i\vQ_1\times\vQ_2}\sum_{m} \rho_{m_1m}(\vq_2)\chi_{mm_2}(\vq_1)
\label{id2a}\eea
Note that the cyclotron parts of $\rho_e$ and $\rho_v$ {\it do not
commute}, even though the entire operators, including the CF-guiding
center parts, do.

\end{document}